\documentclass[12pt,a4paper]{revtex4}
\usepackage{amsmath,amssymb,graphics,epsfig,subfigure}
\usepackage{color}
\usepackage[colorlinks,linkcolor=black,anchorcolor=black,citecolor=green]{hyperref}
\usepackage{geometry}
\makeatletter

\newcommand{\Rmnum}[1]{\expandafter\@slowromancap\romannumeral #1@}
\makeatother

\begin{document}

\thispagestyle{empty}

\begin{center}

\title{The correspondence between thermodynamic curvature and isoperimetric theorem from ultraspinning black hole}

\author{Zhen-Ming Xu$^{}$\footnote{E-mail: xuzhenm@nwu.edu.cn}}

\affiliation{ $^{2}$Institute of Modern Physics, Northwest University, Xi'an 710127, China\\
$^{1}$School of Physics, Northwest University, Xi'an 710127, China\\
$^{3}$Shaanxi Key Laboratory for Theoretical Physics Frontiers, Xi'an 710127, China\\
$^{4}$Peng Huanwu Center for Fundamental Theory, Xian 710127, China}

\begin{abstract}
In this paper, a preliminary correspondence between the thermodynamic curvature and the isoperimetric theorem is established from a $4$-dimensional ultraspinning black hole. We find that the thermodynamic curvature of ultraspinning black hole is negative which means the ultraspinning black hole is likely to present an attractive between its molecules phenomenologically if we accept the analogical observation that the thermodynamic curvature reflects the interaction between molecules in a black hole system. Meanwhile we obtain a general conclusion that the thermodynamic curvature of the extreme black hole of the super-entropic black hole has a (positive or negative) remnant approximately proportional to the reciprocal of entropy of the black hole.
\end{abstract}

\maketitle
\end{center}

\section{Introduction}
A very interesting and challenging problem in black hole thermodynamics is the volume of black hole. Although there are various versions of black hole volume discussion \cite{Parikh2006,Grumiller2006,Ballik2010,Ballik2013,MacDonald2014,Brenna2015,Christodoulou2015,Dolan2011a,Kubiznak2017,Dolan2011b}, there is no unified description yet. In the problem of understanding the volume of black holes, especially in AdS black holes, the application of isoperimetric theorem deepens our mathematical understanding of black hole thermodynamics insofar as it places a constraint on the thermodynamic volume and entropy of an AdS (or dS) black hole \cite{Cvetic2011,Dolan2013}. Isoperimetric theorem is an ancient mathematical problem, which simply means that in a simple closed curve of a given length on a plane, the area around the circumference is the largest. With the proposal of black hole area entropy (in the natural unit system, $S=A/4$, where $S$ is the entropy of the black hole and $A$ is the area of the event horizon) \cite{Bekenstein1973,Bardeen1973} and the introduction of extended phase space \cite{Kastor2009}, Cveti\v{c}, Gibbons, Kubiz\v{n}\'{a}k, and Pope creatively applied the theorem to AdS black hole system and conjectured that in general for any $d$-dimensional asymptotic AdS black hole, its thermodynamic volume $V$ and entropy $S$ satisfy the reverse isoperimetric inequality \cite{Cvetic2011},
\begin{eqnarray}\label{ratio}
\mathcal{R}=\left(\frac{(d-1)V}{\omega_{d-2}}\right)^{\frac{1}{d-1}}\left(\frac{\omega_{d-2}}{4S}\right)^{\frac{1}{d-2}}\geq 1,
\end{eqnarray}
where $\omega_n=2\pi^{(n+1)/2}/\Gamma\left[(n+1)/2\right]$ is the standard volume of the round unit sphere, and the equality is attained for the (charged) Schwarzschild-AdS black hole. Physically, the above isoperimetric ratio indicates that the entropy of black holes is maximized for the (charged) Schwarzschild-AdS black hole at a given thermodynamic volume. Up to now, the ratio has been verified for a variety of black holes with the horizon of spherical topology and black rings with the horizon of toroidal topology \cite{Altamirano2014}. The black hole, which violates the reverse isoperimetric inequality, i.e., $\mathcal{R}<1$, is called a super-entropic black hole \cite{Mann2018}. To date, there are only two known super-entropic black holes. One is $(2+1)$-dimensional charged Banados-Teitelboim-Zanelli (BTZ) black hole which is the simplest \cite{Frassino2015,Johnson2019a,Johnson2019b,Mo2017,Xu2020a}. Another important super-entropic black hole is a kind of ultraspinning black hole \cite{Hennigar2015a,Hennigar2015b,Appels2019}.

Now turn to another important concept, thermodynamic curvature. It is now the most important physical quantity in studying the micro-mechanism of black holes from the axioms of thermodynamics phenomenologically. Its theoretical basis is mainly based on the thermodynamics geometry, which is mainly to use the Hessian matrix structure to represent the thermodynamic fluctuation theory \cite{Ruppeiner1995}. Hitherto without an underlying theory of quantum gravity, the exploration on the microscopic structure of black holes is bound to some speculative assumptions. Owing to the well-established black hole thermodynamics, as an analogy analysis and a primary description, it can be said that the thermodynamic geometry should yet be regarded as probe kits to phenomenologically or qualitatively extract certain information about interactions of black holes. In this scene, one can regard that an empirical observation in ordinary thermodynamics that negative (positive) thermodynamic curvature is associated with attractive (repulsive) microscopic interactions, is also applicable to black hole systems \cite{Ruppeiner2014}. Based on this empirical analogy analysis, the primary microscopic information of the BTZ black hole, (charged) Schwarzschild (-AdS) black hole, Gauss-Bonnet (-AdS) black hole, higher dimensional black holes and other black holes are explored \cite{Ruppeiner2008,Wei2015,Wei2019a,Wei2019b,Wei2019c,Miao2018a,Miao2018b,Miao2019a,Miao2019b,Aman2003,Mirza2007,Dehyadegari2017,Cai1999,Zhang2015a,
Zhang2015b,Liu2010,Xu2019a,Xu2019b,Niu2012,Wang2019,Ghosh2019,Bhattacharya2017,Chen2019,Guo2019,Mansoori2014,Mansoori2015,Sarkar2006,Quevedo2009,Akbar2011,Mohammadzadeh2018,Ghosh2020,Xu2020b}.

In this paper, we shall calculate the thermodynamic curvature of $4$-dimensional ultraspinning black hole and explore the correspondence between thermodynamic curvature and isoperimetric theorem of super-entropic black hole. First, the thermodynamic curvature of ultraspinning black hole has never been analyzed, so we want to fill this gap. Second, the isoperimetric ratio~(\ref{ratio}) has been simply an observation made in the literature, but no physical reason has been given for the bound. Hence we want to try to understand this isoperimetric ratio from the point of view of thermodynamics geometry. Third, in our previous work \cite{Xu2020a} about the thermodynamic curvature of $(2+1)$-dimensional charged BTZ black hole, we give a preliminary conjecture that {\em when the isoperimetric ratio is saturated ($\mathcal{R}=1$), the thermodynamic curvature of an extreme black hole tends to be infinity while for super-entropic black holes ($\mathcal{R}<1$), the thermodynamic curvature of the extreme black hole goes to a finite value}. In present paper, through the analysis of the thermodynamic curvature of the only second super-entropic black hole, we want to verify and perfect the previous conjecture and establish a new correspondence, that is, the correspondence of thermodynamics curvature and isoperimetric theorem of AdS black holes.

\section{Thermodynamic properties of ultraspinning black hole} \label{sec2}
We start to demonstrate this procedure with the $4$-dimensional Kerr-AdS black hole and write its metric in the standard Boyer-Lindquist form \cite{Dolan2011a,Hennigar2015a}
\begin{eqnarray}
ds^2=-\frac{\Delta_a}{\Sigma_a}\left[dt-\frac{a\sin^2 \theta}{\Xi}d\phi\right]^2+\frac{\Sigma_a}{\Delta_a}dr^2+\frac{\Sigma_a}{\Pi}d\theta^2+\frac{\Pi\sin^2\theta}{\Sigma_a}\left[adt-\frac{r^2+a^2}{\Xi}d\phi\right]^2
\end{eqnarray}
where
\begin{eqnarray}
\Sigma_a &=& r^2+a^2\cos^2\theta, \quad \Xi=1-\frac{a^2}{l^2}, \quad \Pi=1-\frac{a^2}{l^2}\cos^2\theta,\nonumber\\
\Delta_a &=& (r^2+a^2)\left(1+\frac{r^2}{l^2}\right)-2mr,
\end{eqnarray}
here $m$ is related to black hole mass, $l$ is the AdS radius which is connected with the negative cosmological constant $\Lambda$ via $\Lambda=-1/l^2$ and $a$ is  rotation parameter.

To avoid a singular metric in limit $a\rightarrow l$, Refs.~\cite{Hennigar2015a,Hennigar2015b} define a new azimuthal coordinate $\psi=\phi/\Xi$ and identify it with period $2\pi/\Xi$ to prevent a conical singularity. After these coordinate transformations and then taking the limit $a\rightarrow l$, one can get the metric of the ultraspinning black hole~\cite{Hennigar2015a,Hennigar2015b}
\begin{eqnarray}
ds^2=-\frac{\Delta}{\Sigma}\left[dt-l\sin^2\theta d\phi\right]^2+\frac{\Sigma}{\Delta}dr^2+\frac{\Sigma}{\sin^2\theta}d\theta^2+\frac{\sin^4\theta}{\Sigma}\left[ldt-(r^2+l^2)d\phi\right]^2
\end{eqnarray}
where
\begin{eqnarray}
\Sigma=r^2+l^2\cos^2\theta, \quad \Delta=\left(l+\frac{r^2}{l}\right)^2-2mr,
\end{eqnarray}
and the horizon $r_h$ defined by $\Delta(r_h)=0$. In addition, due to the new azimuthal coordinate $\psi$ is noncompact, Refs.~\cite{Hennigar2015a,Hennigar2015b} choose to compactify by requiring that $\psi\sim\psi+\mu$ with a dimensionless parameter $\mu$. For this black hole, in order to make the horizon exist, the mass of the black hole is required to have a minimum, that is, an extreme black hole,
\begin{eqnarray}\label{oex}
m\geq m_0=\frac{8}{3\sqrt{3}}l, \qquad r_0=\frac{l}{\sqrt{3}}.
\end{eqnarray}
Correspondingly, the first law of ultraspinning black hole thermodynamics is~\cite{Hennigar2015a,Hennigar2015b}
\begin{equation}\label{olaw}
dM=TdS+VdP+\Omega dJ,
\end{equation}
where the basic thermodynamic properties, i.e., enthalpy $M$, temperature $T$, entropy $S$, thermodynamic pressure $P$, thermodynamic volume $V$, angular momentum $J$ and angular velocity $\Omega$, of ultraspinning black hole associated with horizon radius $r_h$ are~\cite{Hennigar2015a,Hennigar2015b}
\begin{eqnarray}\label{properties}
M&=&\frac{\mu m}{2\pi}, \quad J=Ml, \quad \Omega=\frac{l}{r_h^2+l^2},\nonumber\\
S&=&\frac{\mu(r_h^2+l^2)}{2}, \quad T=\frac{1}{4\pi r_h}\left(\frac{3r_h^2}{l^2}-1\right),\nonumber\\
P&=&\frac{3}{8\pi l^2}, \quad V=\frac{2\mu r_h (r_h^2+l^2)}{3}.
\end{eqnarray}

Meanwhile authors in Refs.~\cite{Hennigar2015a,Hennigar2015b} find the above ultraspinning black hole is super-entropic, i.e., the relation between the entropy $S$ and thermodynamic volume $V$ in Eq.~(\ref{properties}) violates the reverse isoperimetric inequality~(\ref{ratio}).

We notice that the above first law~(\ref{olaw}) is mathematically problematic, like as the Maxwell relation $(\partial T/\partial P)_{_{S,J}}\neq (\partial V/\partial S)_{_{P,J}}$. Because angular momentum $J=Ml$ (it's also known in the Ref.~\cite{Hennigar2015a} as chirality condition), it renders the enthalpy $M$ of a black hole just a function of entropy $S$ and pressure $P$. Hence we need to find a more suitable expression of the first law and the derived expressions of temperature and volume. By inserting the chirality condition into the Eq.~(\ref{olaw}), we can get the {\em right} form of the first law of ultraspinning black hole
\begin{equation}\label{law}
dM=\tilde{T}dS+\tilde{V}dP,
\end{equation}
where
\begin{eqnarray}
\tilde{T}=\frac{r_h^2+l^2}{4\pi r_h}\left(\frac{3}{l^2}-\frac{1}{r_h^2}\right),\label{temperature}
\end{eqnarray}
and
\begin{eqnarray}
\tilde{V}=\frac{\mu l^2(r_h^2+l^2)^2}{4r_h}\left(\frac{2}{l^2}-\frac{1}{r_h^2}\right). \label{volume}
\end{eqnarray}
Of course, naturally, we can verify the Maxwell relation $(\partial \tilde{T}/\partial P)_{_{S}}=(\partial \tilde{V}/\partial S)_{_{P}}$. Meanwhile we can write the corresponding Smarr relation
\begin{equation}
M=2\tilde{T}S-2\tilde{V}P,
\end{equation}
which can also be derived from a scaling (dimensional) argument~\cite{Kubiznak2012}. Next let's check whether the ultraspinning black hole is still super-entropic in our new thermodynamic framework. Keeping in mind that the space is compactified due to $\psi\sim\psi+\mu$, we have $\omega_2=2\mu$~\cite{Hennigar2015a}. For convenience, we set a dimensionless parameter $x=l^2/r_h^2$. Consequently, the isoperimetric ratio reads
\begin{eqnarray}
\mathcal{R}=(1+x)^{1/6}\left(1-\frac{x}{2}\right)^{1/3}.
\end{eqnarray}

Now let's analyze the situation of the extreme black hole in our new thermodynamic framework.
\begin{itemize}
  \item For the black hole thermodynamic system, the temperature and thermodynamic volume of the system should be non-negative (we mainly focus on these two physical quantities and the others are positive). For the case of negative temperature and negative thermodynamic volume, this is beyond the scope of this paper, so we have to exclude this situation. Especially for negative thermodynamic volume, it is not well defined in thermodynamics.
  \item For the ultraspinning black hole, the original extreme black hole corresponds to Eq.~(\ref{oex}). There is a lower bound for the mass of the black hole. In short, the original black hole satisfies the condition $0 \leq x \leq 3$. Under this condition, the temperature and thermodynamic volume are not negative, and the extreme black hole is at $x=3$. But unfortunately, as mentioned earlier, the first law of thermodynamics Eq.~(\ref{olaw}) for the black hole is mathematically problematic.
  \item In our new thermodynamic framework, see Eqs.~(\ref{law}),~(\ref{temperature}), and~(\ref{volume}), we guarantee the {\em right} form of the first law of thermodynamics by introducing new expressions of black hole temperature and thermodynamic volume. In order to ensure the non-negativity of these two thermodynamics quantities, we must require $0 \leq x \leq 2$. Under this new condition, the first law of thermodynamics of the ultraspinning black hole is mathematically reasonable, but the cost is to change the original extreme configuration of the black hole. Specifically, the new extreme black hole is at $x=2$ or corresponds to the new lower bound
      \begin{eqnarray}
      m\geq \tilde{m}_0=\frac{9}{4\sqrt{2}}l, \qquad \tilde{r}_0=\frac{l}{\sqrt{2}}.
     \end{eqnarray}
     This is different from the original extreme black hole structure Eq.~(\ref{oex}).
\end{itemize}

At $0< x \leq 2$, we can easily prove that $\mathcal{R}\leq 1$, which implies that the ultraspinning black hole is still super-entropic in our new thermodynamic framework. When the value of $x$ exceeds $2$, the thermodynamic volume of black hole becomes negative, and the isoperimetric ratio is no longer applicable, so it is impossible to determine whether the ultraspinning black hole is super-entropic or not.

\section{Thermodynamic curvature of ultraspinning black hole}
Now we start to calculate the thermodynamic curvature of the ultraspinning black hole, so as to verify the corresponding relationship proposed by Ref.~\cite{Xu2020a} between the thermodynamic curvature and the isoperimetric theorem, and extract the possible microscopic information of the ultraspinning black hole completely from a thermodynamic point of view.

Considering an isolated thermodynamic system with entropy $S$ in equilibrium, the author Ruppeiner in Refs.~\cite{Ruppeiner1995,Ruppeiner2014,Ruppeiner2008} divided it into a small subsystem $S_B$ and a large subsystem $S_E$ with requirement of $S_B \ll S_E \sim S$. We have known that in equilibrium state, the isolated thermodynamic system has a local maximum entropy $S_0$ at $x_0^\mu$. Hence at the vicinity of the local maximum, we can expand the entropy $S$ of the system to a series form about the equilibrium state
\begin{equation}
    S=S_0+\frac{\partial S_B}{\partial x_B^\mu}\Delta x^\mu_B
          +\frac{\partial S_E}{\partial x_E^\mu}\Delta x^\mu_E
      +\frac{1}{2}\frac{\partial^2 S_B}{\partial x_B^\mu \partial x_B^\nu}\Delta x^\mu_B \Delta x^\nu_B
       +\frac{1}{2}\frac{\partial^2 S_E}{\partial x_E^\mu \partial x_E^\nu}\Delta x^\mu_E \Delta x^\nu_E
       +\cdots,
\end{equation}
where $x^\mu$ stand for some independent thermodynamic variables. Due the conservation of the entropy of the equilibrium isolated system and the condition $S_B \ll S_E \sim S$, the above formula approximately becomes
\begin{equation}
\Delta S =S_0-S \approx -\frac{1}{2}\frac{\partial^2 S_B}{\partial x_B^\mu \partial x_B^\nu}\Delta x^\mu_B \Delta x^\nu_B,
\end{equation}
where the so-called Ruppeiner metric is (here we omit subscript $B$)
\begin{equation}\label{rmetric}
\Delta l^2=-\frac{\partial^2 S}{\partial x^\mu \partial x^\nu}\Delta x^\mu \Delta x^\nu=g^S_{\mu\nu}\Delta x^\mu \Delta x^\nu.
\end{equation}

Now focus on the system of the ultraspinning black hole and its surrounding infinite environment. Black hole itself can be regarded as the small subsystem mentioned above. In the light of the {\em right} form of the first law of thermodynamics Eq.~(\ref{law}), we can get the general form of the Ruppeiner metric for the ultraspinning black holes
\begin{equation}
\Delta l^2=\frac{1}{\tilde{T}}\Delta \tilde{T} \Delta S+\frac{1}{\tilde{T}}\Delta \tilde{V} \Delta P.
\end{equation}

In principle, according to the first law Eq.~(\ref{law}), the phase space of the ultraspinning black hole is $\{\tilde{T}, P, S, \tilde{V}\}$. For the thermodynamics geometry, it is carried out in the space of generalized coordinates, like as $\{S,P\}$, $\{S,\tilde{V}\}$, $\{\tilde{T},\tilde{V}\}$ and $\{\tilde{T},P\}$. There is Legendre transformation between the thermodynamic potential functions corresponding to these coordinate spaces. Hence the thermodynamic curvatures obtained in these coordinate spaces are same. For avoiding the technique complexity, we take the coordinate space $\{S,P\}$ as an example for detailed calculation. The line element of thermodynamic geometry becomes~\cite{Xu2020a,Xu2019a}
\begin{equation}\label{linesp}
\begin{aligned}
\Delta l^2 &=\frac{1}{\tilde{T}}\left(\frac{\partial \tilde{T}}{\partial S}\right)_P \Delta S^2+\frac{2}{\tilde{T}}\left(\frac{\partial \tilde{T}}{\partial P}\right)_S \Delta S \Delta P+\frac{1}{\tilde{T}}\left(\frac{\partial \tilde{V}}{\partial P}\right)_S \Delta P^2\\
&=\frac{1}{\tilde{T}}\frac{\partial^2 M}{\partial X^\mu \partial X^\nu}\Delta X^\mu \Delta X^\nu=g_{\mu\nu}\Delta X^\mu \Delta X^\nu, \quad (\mu, \nu=1,2)
\end{aligned}
\end{equation}
where $(X^1, X^2)=(S, P)$ and in the right part of the second equal sign, we have used the first law of thermodynamics Eq.~(\ref{law}). The above thermodynamic metric $g_{\mu\nu}$ is equivalent to the metric $g^S_{\mu\nu}$ in Eq.~(\ref{rmetric}), but they have different representation forms. The metric $g^S_{\mu\nu}$ is in the entropy representation, while the metric $g_{\mu\nu}$ is in the enthalpy representation. Next according to the specific form of the metric $g_{\mu\nu}$, we start to calculate the thermodynamic curvature, which is the ``thermodynamic analog'' of the geometric curvature in general relativity. By using the Christoffel symbols
$\Gamma^{\alpha}_{\beta\gamma}=g^{\mu\alpha}\left(\partial_{\gamma}g_{\mu\beta}+\partial_{\beta}g_{\mu\gamma}-\partial_{\mu}g_{\beta\gamma}\right)/2$
and the Riemannian curvature tensors
${R^{\alpha}}_{\beta\gamma\delta}=\partial_{\delta}\Gamma^{\alpha}_{\beta\gamma}-\partial_{\gamma}\Gamma^{\alpha}_{\beta\delta}+
\Gamma^{\mu}_{\beta\gamma}\Gamma^{\alpha}_{\mu\delta}-\Gamma^{\mu}_{\beta\delta}\Gamma^{\alpha}_{\mu\gamma}$,
we can obtain the thermodynamic curvature $R_{_{SP}}=g^{\mu\nu}{R^{\xi}}_{\mu\xi\nu}$.

With the help of Eqs.~(\ref{temperature}),~(\ref{volume}) and the expressions of entropy $S$ and thermodynamic pressure $P$ in Eq.~(\ref{properties}), the thermodynamic curvature can be directly read as
\begin{equation}\label{curvature}
R_{_{SP}}=-\frac{x(x+1)[x^2(x-3)(x^2+12)+27x-9]}{2S(x-3)[x^2(x-3)+3x-3]^2}.
\end{equation}

In view of the thermodynamic curvature obtained above, some explanations are made.
\begin{itemize}
  \item For the extreme black hole, i.e., $x=2$, we can observe clearly that thermodynamic curvature is finite negative value $R_{SP}|_{_{\text{extreme}}}=-57/S$.
  \item Due to $0< x \leq 2$, with a little calculation, we can obtain $R_{_{SP}}<0$. We can speculate that the ultraspinning black hole is likely to present a attractive between its molecules phenomenologically or qualitatively.
  \item Look at the original extreme black hole, i.e., $x=3$, you might intuitively get that the thermodynamic curvature tends to be infinite at this time according to Eq.~(\ref{curvature}). In fact, in this case, the basic thermodynamic metric~(\ref{linesp}) is no longer valid, because the first law~(\ref{olaw}) is pathological.
\end{itemize}

At present, the known super-entropic black holes are only (2+1)-dimensional charged BTZ black hole and the ultraspinning black hole. According to our current analysis and the calculation of charged BTZ black hole in the previous paper \cite{Xu2020a}, we have for ultraspinning black hole $R_{SP}|_{_{\text{extreme}}}=-57/S$ and for charged BTZ black hole $R_{SP}|_{_{\text{extreme}}}=1/(3S)$. Hence, a universal relationship is
\begin{equation}\label{excurvature}
R_{SP}|_{_{\text{extreme}}}\propto\frac{1}{S}.
\end{equation}
We know that the reverse isoperimetric inequality physically indicates that at a given thermodynamic volume, the (charged) Schwarzschild-AdS black holes  are maximally entropic. The super-entropic black hole means that the entropy of black hole exceeds the maximal bound. For the (charged) Schwarzschild-AdS black hole, the thermodynamic curvature of the corresponding extreme black hole tends to be infinity, which is verified by various simple static black hole solutions of the pure Einstein gravity or higher-derivative generalizations thereof. Therefore, we can have the following corresponding relations:
\begin{itemize}
  \item For the black holes with $\mathcal{R}=1$, the thermodynamic curvature of the corresponding extreme black hole tends to be (positive or negative) infinity.
  \item For the black holes with $\mathcal{R}<1$, the thermodynamic curvature of the corresponding extreme black hole has a (positive or negative) remnant which is approximately proportional to $1/S$.
  \item For the black holes with $\mathcal{R}>1$, the thermodynamic curvature of the corresponding extreme black hole is also (positive or negative) infinity.
\end{itemize}

We note that the last conjecture mentioned above about the extreme behavior of the thermodynamic curvature of the sub-entropic black hole ($\mathcal{R}>1$), needs further verification in the future. At present, we only think that at the case of exceeding the maximum entropy, the thermodynamic curvature of the corresponding extreme black hole has a finite remnant, but at the case of the maximum entropy, it tends to infinity. Naturally, when the entropy of black hole is less than the maximum entropy, the thermodynamic curvature of the corresponding extreme black hole tends to infinity intuitively.

\section{Conclusion and Discussion}
In this paper, we investigate the the thermodynamic curvature of the ultraspinning black hole by introducing the {\em right} form of the first law~(\ref{law}). We find that the ultraspinning black hole is still super-entropic in our new thermodynamic framework, which is consistent with the result obtained in \cite{Hennigar2015a,Hennigar2015b}. Meanwhile the obtained thermodynamic curvature is negative which means the ultraspinning black hole is likely to present a attractive between its molecules phenomenologically or qualitatively if we accept the analogical observation that the thermodynamic curvature reflects the interaction between molecules in a black hole system. Through the analysis of the extreme behavior of the thermodynamic curvature, we can get a general conclusion that the thermodynamic curvature of the extreme black hole of the super-entropic black hole has a (positive or negative) remnant approximately proportional to $1/S$. This is a very interesting result.

In our previous work \cite{Xu2019a}, we analyze the thermodynamic curvature of Schwarzschild black hole and obtain $R_{\text{Schwarzschild}}=\pm 1/S_{\text{Schwarzschild}}$. This one is very similar to what we've got in present paper. Maybe it's a coincidence? Maybe it suggests that the excess entropy in the super-entropic black hole comes from the Schwarzschild black hole? This unexpected problem needs further analysis and discussion.

Furthermore, we need to in the future confirm the conjecture about the sub-entropic black hole, such as the Kerr-AdS black hole \cite{Cvetic2011,Johnson2019c}, STU black holes \cite{Johnson2019c,Caceres2015}, Taub-NUT/Bolt black hole \cite{Johnson2014}, generalized exotic BTZ black hole \cite{Johnson2019b}, noncommutative black hole \cite{Miao2017} and accelerating black holes\cite{Appels2016}. The verification of this conjecture will help us to improve the correspondence between the thermodynamic curvature and the isoperimetric theorem, which is a very meaningful research content.

\section*{Acknowledgments}
The financial support from National Natural Science Foundation of China (Grant Nos. 11947208 and 11947301) is gratefully acknowledged. This research is also supported by The Double First-class University Construction Project of Northwest University. The author would like to thank the anonymous reviewers for the helpful comments that indeed greatly improve this work.

\end{document}